\documentclass[nofootinbib,aps,twocolumn,superscriptaddress]{revtex4}

\usepackage{natbib}
\setcitestyle{square, comma, numbers,sort&compress, super}
\RequirePackage{graphicx,amsmath}
\RequirePackage{mathptmx}    
\RequirePackage{flushend}
\RequirePackage{lineno}
\usepackage{doi}
\usepackage[]{units}
\usepackage{xspace}
\usepackage{pbox}
\usepackage{multirow}
\usepackage{lineno}
\usepackage{xcolor}
\usepackage{morefloats}

\def\mymathhyphen{{\hbox{-}}}

\newcommand{\Na}{$^{22}$Na}
\newcommand{\K}{$^{40}$K}
\newcommand{\roi}{\ensuremath{\rm[1\mymathhyphen6]}}
\newcommand{\sroi}{\ensuremath{\rm[2\mymathhyphen6]}}
\newcommand{\anais}{\ensuremath{\rm ANAIS\mymathhyphen112}}

\begin{document}

\title{First results on dark matter annual modulation from ANAIS-112 experiment}

\author{J.~Amar{\'e}}
\author{S.~Cebri{\'an}}
\author{I.~Coarasa}
\affiliation{Laboratorio de F\'isica Nuclear y Astropart\'iculas, Universidad de Zaragoza, C/ Pedro Cerbuna 12, 50009 Zaragoza, Spain}
\affiliation{Laboratorio Subterr{\'a}neo de Canfranc, Paseo de los Ayerbe s.n., 22880 Canfranc Estaci{\'o}n, Huesca, Spain}
\author{C.~Cuesta}
\affiliation{Laboratorio de F\'isica Nuclear y Astropart\'iculas, Universidad de Zaragoza, C/ Pedro Cerbuna 12, 50009 Zaragoza, Spain}
\affiliation{Present Address: Centro de Investigaciones Energ\'eticas, Medioambientales y Tecnol\'ogicas, CIEMAT, 28040, Madrid, Spain}
\author{E.~Garc\'{\i}a}
\affiliation{Laboratorio de F\'isica Nuclear y Astropart\'iculas, Universidad de Zaragoza, C/ Pedro Cerbuna 12, 50009 Zaragoza, Spain}
\affiliation{Laboratorio Subterr{\'a}neo de Canfranc, Paseo de los Ayerbe s.n., 22880 Canfranc Estaci{\'o}n, Huesca, Spain}
\author{M.~Mart\'{\i}nez}
\affiliation{Laboratorio de F\'isica Nuclear y Astropart\'iculas, Universidad de Zaragoza, C/ Pedro Cerbuna 12, 50009 Zaragoza, Spain}
\affiliation{Laboratorio Subterr{\'a}neo de Canfranc, Paseo de los Ayerbe s.n., 22880 Canfranc Estaci{\'o}n, Huesca, Spain}
\affiliation{Fundaci\'on ARAID, Av. de Ranillas 1D,  50018 Zaragoza, Spain}
\author{M.A. Oliv{\'a}n}
\affiliation{Laboratorio de F\'isica Nuclear y Astropart\'iculas, Universidad de Zaragoza, C/ Pedro Cerbuna 12, 50009 Zaragoza, Spain}
\affiliation{Present Address: Fundaci{\'o}n CIRCE, 50018, Zaragoza, Spain}
\author{Y.~Ortigoza}
\author{A.~Ortiz~de~Sol{\'o}rzano}
\author{J.~Puimed{\'o}n}
\author{A.~Salinas}
\author{M.L.~Sarsa\footnote{Corresponding author: mlsarsa@unizar.es}}
\author{P.~Villar}
\author{J.A.~Villar\footnote{Deceased}}
\affiliation{Laboratorio de F\'isica Nuclear y Astropart\'iculas, Universidad de Zaragoza, C/ Pedro Cerbuna 12, 50009 Zaragoza, Spain}
\affiliation{Laboratorio Subterr{\'a}neo de Canfranc, Paseo de los Ayerbe s.n., 22880 Canfranc Estaci{\'o}n, Huesca, Spain}

\begin{abstract}
ANAIS is a direct detection dark matter experiment aiming at the testing of the DAMA/LIBRA annual modulation result, which standing for about two decades has 
neither been confirmed nor ruled out by any other experiment in a model independent way. 
{\anais}, consisting of 112.5~kg of sodium iodide crystals, is taking data at the 
Canfranc Underground Laboratory, Spain, since August 2017. 
This letter presents the annual modulation analysis of 1.5 years of data, 
amounting to 157.55~kg$\times$y. We focus on the model 
independent analysis searching for modulation and the validation of our 
sensitivity prospects. 
{\anais} data are consistent with the null hypothesis (p-values of 0.65 and 0.16 for 
{\sroi} and {\roi}~keV energy regions, respectively). 
The best fits for the modulation hypothesis are consistent with the absence of modulation 
($S_m=-0.0044\pm0.0058$~cpd/kg/keV and $-0.0015\pm0.0063$~cpd/kg/keV, respectively). They are in agreement with our estimated sensitivity for the accumulated exposure, supporting our projected goal of reaching a 3$\sigma$ sensitivity to the DAMA/LIBRA result in 5 years of data taking.
\end{abstract}

\maketitle
\vline



There is overwhelming evidence from cosmological and astrophysical observations supporting that a large fraction of the Universe energy-mass budget is not explained in the framework of the standard model of the particle physics, assuming the cosmological standard model~\cite{Bertone:2016nfn}. The solution to the dark matter (DM) / dark energy (DE) puzzle is probably of a complex nature. In one of the preferred hypothetical scenarios, DM can be understood as a new non-zero-mass particle having a very low interaction probability with baryonic matter. Although proposed candidates span about 45 orders of magnitude in mass, and 60 in cross-section, axions and Weakly Interacting Massive Particles (WIMPs) are among the better motivated~\cite{Baudis:2016qwx}. 
Experimental effort devoted to unraveling the nature of the DM particles has been spent, either by direct~\cite{Liu:2017drf}, indirect~\cite{Conrad:2017pms} or accelerator searches~\cite{Buchmueller:2017qhf}, which are complementary to each other. Only one experiment, DAMA/LIBRA~\cite{Bernabei:2008yi, Bernabei:2013xsa, Bernabei:2018yyw}, has provided a long-standing positive result: the observation of a highly statistically significant annual modulation in the detection rate, compatible with that expected for galactic halo dark matter particles. 
This result has neither been reproduced by any other experiment, nor ruled out in a model independent way. Compatibility among the different experimental results in most conventional WIMP-DM scenarios is actually disfavored~\cite{Adhikari:2018ljm,Kobayashi:2018jky,Akerib:2018zoq,Abe:2018mxq,Aprile:2017yea,Savage:2009mk,Aprile:2015ade,Herrero-Garcia:2015kga,Baum:2018ekm,Kang:2018qvz,Herrero-Garcia:2018mky}. 
Then, a similar annual modulation search using the same target is mandatory to shed light on the DAMA/LIBRA conundrum, which is the goal of the ANAIS (Annual modulation with NaI Scintillators) experiment. 
Other efforts sharing ANAIS goal in the international dark matter community are the COSINE-100 experiment, taking data also in dark matter mode at Yang-Yang Underground Laboratory, South Korea~\cite{Kim:2018wcl,Adhikari:2017esn,Adhikari:2018ljm}; and in longer-term, SABRE, aiming at installing twin detectors in Australia and Italy \cite{Tomei:2017rkg}, and COSINUS, developing cryogenic detectors based on NaI~\cite{Gutlein:2017iuw}. 
\par
An annual modulation in the dark matter interaction rate is expected by the revolution of the Earth around the Sun, which distorts the DM particle velocity distribution function as seen by the detector, typically assumed Maxwellian boosted by the Sun velocity~\cite{Freese:1987wu,Freese:2012xd}.
The effect is present unless the DM halo is co-rotating with the Solar System. However, it is strongly dependent on the specific halo model, both in amplitude and in phase. It is natural to assume that the 
Sun is moving through a locally isotropic DM halo, with the Earth orbiting aside. 
Consequently, searches are performed for a modulation of DM-like events with a one year period
and a well defined phase.
On the other hand, preferably an annual modulation analysis should not assume a priori neither the period of the modulation nor the phase, but it should derive them from the data. A full and consistent analysis requires then several years of measurement in very stable conditions. This is the long-term goal of our experiment. 
\par
{\anais}, consisting of 112.5~kg of NaI(Tl) detectors, was installed in 2017 at the Canfranc Underground Laboratory, LSC, in Spain.
The {\anais} set-up undergoes a different residual cosmic ray flux and environmental conditions than DAMA/LIBRA (800~m versus 1400~m rock overburden, for instance). 
Consequently, the potential confirmation of a modulation with same phase and amplitude would be very difficultly explained as an effect of backgrounds or systematics. 
{\anais} experimental details have been recently reported, as well as the performance of the first year's operation~\cite{Amare:2018sxx}, analysis of backgrounds~\cite{Amare:2018ndh}, and sensitivity prospects for a five-years operation~\cite{Coarasa:2018qzs}; here we just briefly summarize the most relevant features of the experimental apparatus. 
\par
{\anais} uses nine NaI(Tl) modules produced by Alpha Spectra Inc. in Colorado (US)\footnote{http://www.alphaspectra.com/}. 
These modules have been manufactured from 2012 to 2017, and shipped to Spain avoiding air travel in order to prevent cosmogenic activation of the module materials. Each crystal is cylindrical (4.75" diameter and 11.75" length), with a mass of 12.5~kg, and it is housed in 
OFE (Oxygen Free Electronic) copper. This encapsulation has a Mylar window allowing low energy calibration using external gamma sources. 
It incorporates two quartz optical windows to couple the photomultipliers (PMTs). 
All PMT units, as well as all relevant materials used in the building of the detectors, have been screened for radiopurity using HPGe detectors in the low background facilities at LSC.
Their contribution to the experiment background has been estimated~\cite{Amare:2016rbf,Amare:2018ndh} and
included in our background model (see below).
We would like to emphasize the outstanding light collection measured for the nine modules, at the level of 15~photoelectrons (phe) per keV~\cite{Amare:2018sxx,Olivan:2017akd}. 
{\anais} is calibrated every two weeks using external $^{109}$Cd sources: all the nine modules are simultaneously calibrated using a multi-source system which minimizes down time periods. 
Background events from the decay of {\K} and {\Na} in the crystal bulk, associated to 3.2 and 0.9~keV energy depositions, and selected by coincidence with an energy deposition in a second module around 1461 and 1275~keV, respectively, are also used to improve the accuracy of the calibration down to the energy threshold~\cite{Amare:2018sxx}. 
\par
The {\anais} shielding consists of 10~cm of archaeological lead, 20~cm of low activity lead, an anti-radon box (continuously flushed with radon-free nitrogen gas), an active muon veto system made up of 16 plastic scintillators designed to cover top and sides of the whole ANAIS set-up and 40~cm of neutron moderator (a combination of water tanks and polyethylene blocks). 
In the design of the muon veto system we follow
a tagging strategy instead of a hardware vetoing. 
The goal is twofold: on the one hand, to discard events in the NaI(Tl) crystals coincident with muon veto triggers.
On the other hand, to  analyse eventual correlations between muon hits in the plastic scintillators and events in the NaI(Tl) crystals, specially in the region 
of interest (ROI), {\roi}~keV\footnote{Energy will be shown in electron equivalent units throughout this letter}. 
\par 
The {\anais} electronic chain and data acquisition system (DAQ) have been described in Refs.~\cite{MAThesis,Amare:2018sxx}. Each PMT charge signal is independently processed and  
divided into: (1) a trigger signal; (2) a low energy (LE) signal that goes to the digitizers which sample the 
waveforms at 2~Gs/s with high resolution (14 bits); and (3) a high energy (HE) signal, conveniently attenuated. 
The trigger of each PMT signal is done at phe level, while the single module trigger is done by the coincidence (logical AND) of the two PMT triggers in a 200~ns window. The global trigger is the logical OR of the nine modules trigger signals. Trigger efficiency is close to 100\% down to the analysis threshold established at 1~keV~\cite{Amare:2018sxx}.
\par 
{\anais} started taking data in the DM mode on August, 3$^{rd}$, 2017. It has accumulated almost 1.5 years of data-taking time in quite stable conditions by February, 12$^{th}$, 2019. Total live time available for the annual modulation analysis is 527.08 days: 341.72 (first year) + 185.36 (half second year). This implies a live time of 94.5\% (95.2\%), dead time of 2.9\% (2.2\%), and down time of 2.6\% (2.6\%) for the first (second) year of data taking, respectively. The down time is mainly due to the periodical calibration runs carried out using low energy gamma sources.
We remove events arriving less than one second after the last muon veto trigger, correcting the total live time by subtracting 
one second per muon veto trigger, so the live time used for the annual modulation analysis which follows is 511.16 days.
\par
A blind analysis strategy in three levels has been followed: first, we calculate pulse parameters, the time since the last muon veto, and we apply the peak-finding algorithm to identify individual phe in low energy pulses; second, we calibrate the energy response of every detector at LE and HE~\cite{Amare:2018sxx}; third, we optimize the pulse shape cuts and calculate their efficiency, being the LE variable hidden for events corresponding to single hits (M1 events). Only 10\% of the data were unblinded for general background assessment and fine tuning of procedures~\cite{Amare:2018sxx}; those data were randomly chosen (34~days amounting to 32.9~days live time) and time evolution was kept hidden until the data unblinding, presented in this letter. 
\par 
Calibration procedures and efficiency corrections applied in the following have been derived as done in Ref.~\cite{Amare:2018sxx} for the first year, 
continuing with same procedure in the half second year added in this analysis. 
Events in the ROI are selected, after energy calibration, by imposing the following criteria: single hit events (M1); 
a pulse shape cut combining the fraction of the pulse area in [100-600]~ns after the event trigger, defined following~\cite{2008NIMPA.592..297B}, and the logarithm of the mean time of the distribution of the individual phe arrival times in the digitized window~\cite{Kim:2018wcl}; 
events having an asymmetric light sharing between the two PMT signals are removed by imposing a cut on the 
number of peaks identified in each PMT (asymmetry cut).
The total detection efficiency, $\epsilon(E,d)$, calculated independently for every detector $d$ as a function of the energy, $E$, can be written~\cite{Amare:2018sxx} as
\begin{equation}
\epsilon(E,d)=\epsilon_{trg}(E,d)\times\epsilon_{PSA}(E,d)\times\epsilon_{asy}(E,d),
\end{equation}
where the trigger efficiency $\epsilon_{trg}(E,d)$ is calculated from Monte Carlo (MC) simulations, and 
the efficiencies of the pulse shape, $\epsilon_{PSA}(E,d)$, and asymmetry, $\epsilon_{asy}(E,d)$, cuts are evaluated 
from the 3.2 and 0.9~keV events selected by the coincidence with the high energy gammas following {\K} and {\Na} decays, and $^{109}$Cd calibration events, respectively, accumulated for all the analysed exposure. Total detection efficiency ranges from 0.15 to 0.35 at 1~keV, depending on
the detector, increases up to 0.8 at 2~keV and is nearly 1 at 4~keV
for all the modules. Statistical errors in the total efficiency vary from 3 to 8\% at 1~keV down to 1\% at 6~keV. Comparing 
different methods for the efficiency calculation we have also estimated a systematic uncertainty that amounts up to 20\% at 
1-1.2~keV and is negligible above 1.5~keV. More details can be found in~\cite{Amare:2018sxx}.
\par
The background model for all the nine detectors used in the {\anais} set-up has been developed. It is based on MC simulations using 
the measured activity in external components and
in crystals, including cosmogenic products, quantified in dedicated, independent measurements using different analysis techniques. It provides a good overall description of measured data at all energy ranges above 2~keV
and at different analysis conditions (coincidence or anticoincidence)~\cite{Amare:2018ndh}. In the ROI the background is dominated by the emissions from the crystals themselves, in particular, $^{210}$Pb (32.5\%) and $^3$H (26.5\%) continua, and $^{40}$K (12\%) and $^{22}$Na (2.0\%) peaks are the most significant contributions. Short-lived isotopes activated cosmogenically are still present in the bulk of the last received crystals~\cite{Villar:2018ymt,Amare:2014bea}, contributing as background in the ROI, specially in the [3-5]~keV region. However, from 1 to 2~keV there is a large fraction of our background lacking from explanation~\cite{Amare:2018ndh}. It could have as origin non-bulk scintillation leaking through our event selection criteria. 
\par
\begin{figure}[t]
\centering
\includegraphics[width=0.45\textwidth]{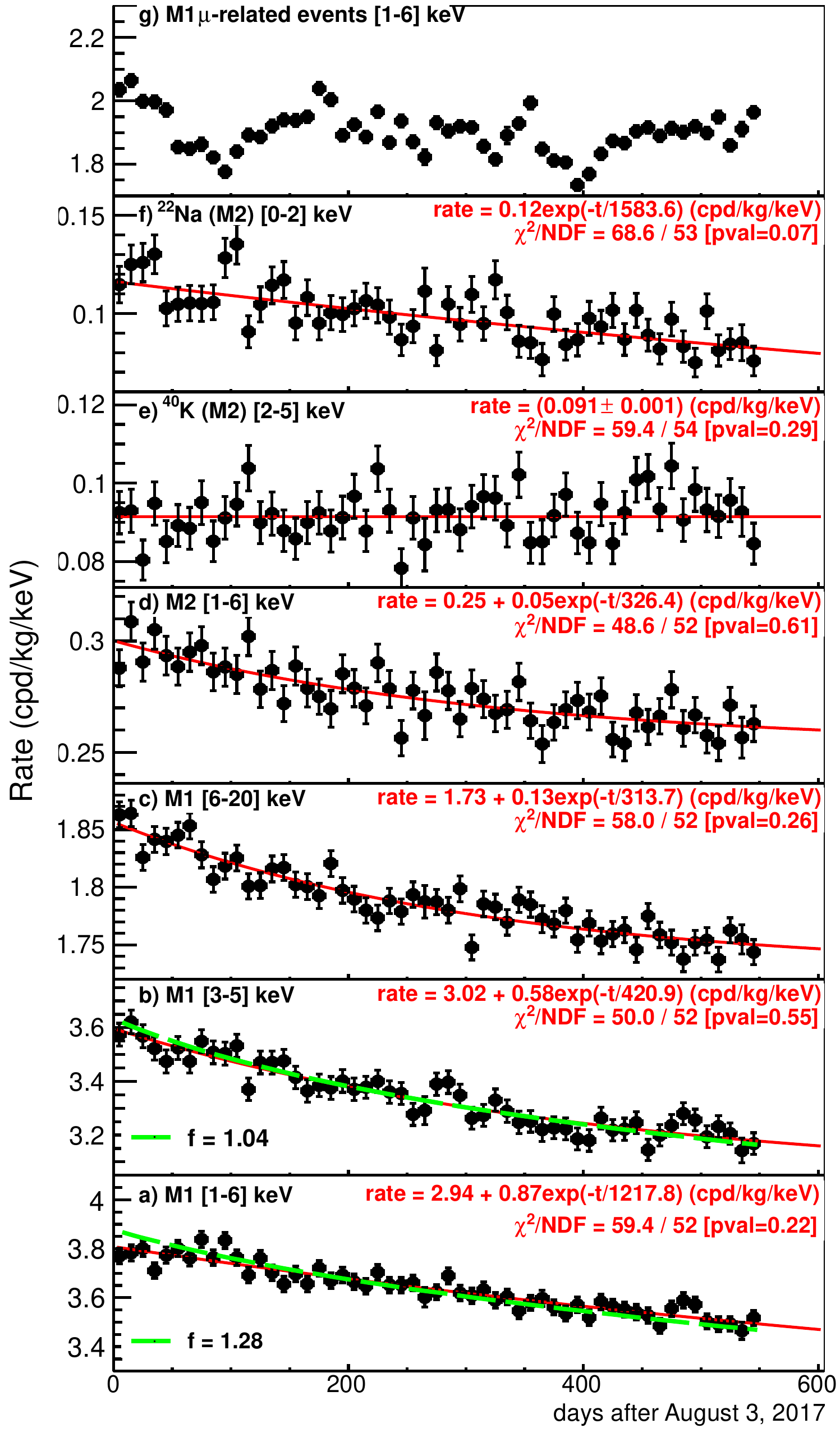}
\caption{Rate of events corresponding to different populations along the first year and a half of {\anais} operation, calculated in ${\rm10\mymathhyphen day}$ binning. Events surviving all the filters are shown in panels a)-f): in energy region {\roi}~keV, a); ${\rm[3\mymathhyphen5]}$~keV, b); ${\rm[6\mymathhyphen20]}$~keV, c); M2 events in energy region {\roi}~keV, d); M2 events in energy region ${\rm[2\mymathhyphen5]}$~keV in coincidence with a high energy gamma in a second detector in the {\K} window, e); and M2 events in energy region ${\rm[0\mymathhyphen2]}$~keV in coincidence with a high energy gamma in a second detector in the {\Na} window, f). Events arriving less than one second after a muon veto trigger before the cuts are shown in panel g). Fits are shown as red solid lines, and corresponding fit parameters and chi-squared and p-values are also given in the plot. Green dashed lines correspond to the background model, normalized according to a factor, $f$, which is also given in the plots.}
\label{fig:mod1}
\end{figure}

The time evolution of the rate of those events surviving all the cuts 
during the first year and a half of {\anais} operation 
is shown for different populations in panels a)-f) of Figure~\ref{fig:mod1}.
Data from all the modules have been added together and corrected by the corresponding efficiencies. 
The two lower panels correspond to two different energy windows in the ROI: {\roi}~keV, a), and [3-5]~keV, b), 
while the upper panels show the evolution of control populations for which no modulation is expected: 
c) [6-20]~keV, d) double-coincidence events (M2) in energy region {\roi}~keV, and coincident events attributed to e) {\K}, and f) {\Na} 
decays in the crystals\footnote{Low energy M2 events having an energy deposition in a second detector in a window around the corresponding high energy gamma}.
The uppermost panel, g), presents the evolution of muon-related low-energy events before the cuts (M1 events in {\roi}~keV region arriving less than 1 second after a muon veto trigger).
As pointed out in Ref.~\cite{Amare:2018sxx}, many low energy events are registered after a muon passage through the detector. They likely originate in the long tail of 
the scintillation pulse produced by a large muon energy deposition, which is able to trigger many times the DAQ system and produce fake low-energy events. This population is clearly fluctuating in time and will be studied in a future work.  

\par
In panels a) and b) of Figure~\ref{fig:mod1} we observe a relevant decrease in the rate at the ROI, which amounts up to 8\% in the analysed period. 
It is caused by cosmogenically activated isotopes and is well reproduced qualitatively in all energy windows by our background model~\cite{Amare:2018ndh}, shown in Figure~\ref{fig:mod1} in green dashed lines. 
It is normalized by a factor, $f$, to be more easily compared with the measured rates. 
Agreement in [3-5]~keV region is also quantitatively very good, $f$=1.04, as the background time evolution in this region is dominated by short-lived cosmogenic isotopes. However, our background model underestimates the rate in {\roi}~keV region ($f$=1.28), because from 1 to 2~keV our background model does not reproduce the measurement~\cite{Amare:2018ndh}, as commented above. 
\par 
We perform a least-squares fit on panels a) to d) to a constant term plus an exponential function, 
to account for the mentioned background reduction, following our background model~\cite{Amare:2018ndh}. 
The $\chi^2/NDF$ and the values of the 
fitted parameters are shown in the figure. 
Good fits are obtained, with p-values larger than 0.20 in all cases.
The {\K} (3.2~keV) and {\Na} (0.9~keV) M2 populations, on the other hand, are modelled by a constant and 
an exponential decay, respectively. We derive consistent results (p-value\,=\,0.22) in the first case, and a reasonable agreement with the {\Na} half-life ($T_{1/2}=3.01\pm0.36$ y, p-value\,=\,0.07) in the second.

In this letter we present our modulation results in two energy regions: {\sroi}~keV and {\roi}~keV, to allow direct 
comparison with the DAMA/LIBRA results. The values of the modulation amplitude observed by 
DAMA/LIBRA, $S_m^{DAMA}$, are $0.0102\pm0.0008$ and $0.0105\pm0.0011$~cpd/kg/keV in the full exposure for {\sroi}~keV and using only phase-2 data for {\roi}~keV energy region, 
respectively~\cite{Bernabei:2018yyw}. 
We expect results derived from {\sroi}~keV to be more robust because our data selection efficiencies strongly go down below 2~keV, increasing the 
risk to be affected by unknown systematics. 
\par

\begin{figure}[hb]
\centering
\includegraphics[width=0.50\textwidth]{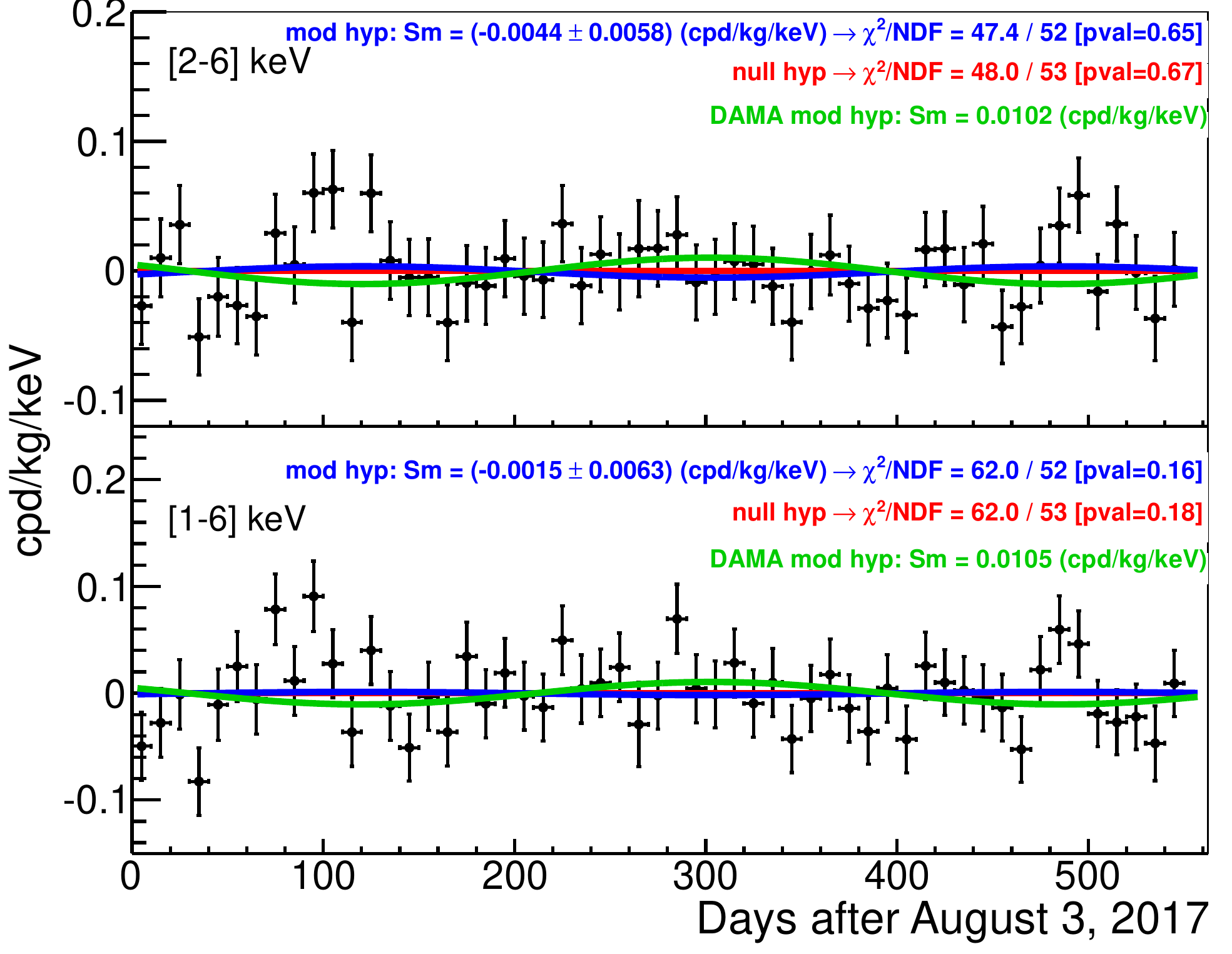}
\caption{{\anais} data 
in the energy windows {\roi}~keV (bottom panel) and {\sroi}~keV (top panel) surviving all the cuts and efficiency corrected~\cite{Amare:2018sxx}. 
Data is displayed after subtracting the constant and exponential functions fitted to Equation~\ref{eq:1}.
Fits are also shown in the same way, both in the modulation 
(3 free parameters) and the null hypothesis (2 free parameters). $\chi^2$ 
and p-values displayed allow the comparison of both hypothesis, and 
DAMA/LIBRA results on modulation amplitude in both energy windows are shown in green~\cite{Bernabei:2018yyw}.}
\label{fig:mod2}
\end{figure}

We evaluate the statistical significance of a possible modulation in our data by a least square method in the time-binned data.
The efficiency-corrected rate of events surviving the cuts in {\roi} and {\sroi}~keV energy regions 
is modelled as
\begin{equation}
R(t) = R_0 + R_1 \cdot exp(-t/\tau) + S_m \cdot cos(\omega \cdot (t + \phi)),
\label{eq:1}
\end{equation}
where $R_0$ and $R_1$ are free parameters and $\tau$ is fixed to the value obtained from our background model in the 
corresponding energy range.
We also fix the period ($\omega=2\pi/365$\,d\,$=0.01721$\,rad\,d$^{-1}$) and the phase ($\phi=-62.2$\,d, corresponding the cosine maximum to June, 2 when taking as time origin August 3), while S$_m$ is fixed to 0 for the null hypothesis and left unconstrained (positive or negative) for the modulation hypothesis. This allows a direct comparison 
with the results from the DAMA/LIBRA analysis with 1 free parameter~\cite{Bernabei:2018yyw}. 
We present the best fit for both hypothesis for ${\rm10\mymathhyphen day}$ time binning in Figure~\ref{fig:mod2}.
In order to highlight the presence or absence of modulation, we plot the data with 
the constant and exponential terms subtracted.
For the sake of comparison, in the plot we show the 
modulation measured by DAMA/LIBRA (green lines).

In both energy regions the null hypothesis is well supported by the $\chi^2$ test, 
with $\chi^2/NDF=48.0/53$
for the {\sroi}~keV (p-value\,=\,0.67) and $\chi^2/NDF=62.0/53$ for the {\roi}~keV 
regions (p-value\,=\,0.18).  
The best fits for the modulation hypothesis
are $S_m=-0.0044\pm0.0058$~cpd/kg/keV and $-0.0015\pm0.0063$~cpd/kg/keV for 
{\sroi}~keV and {\roi}~keV, respectively.
In both cases, p-values are slightly 
lower than those of the null hypothesis (0.65 and 0.16, respectively). 
The best fits are incompatible at 2.5$\sigma$ (1.9$\sigma$) with the DAMA/LIBRA signal.
\par
\begin{figure}[hb]
\centering
\includegraphics[width=0.45\textwidth]{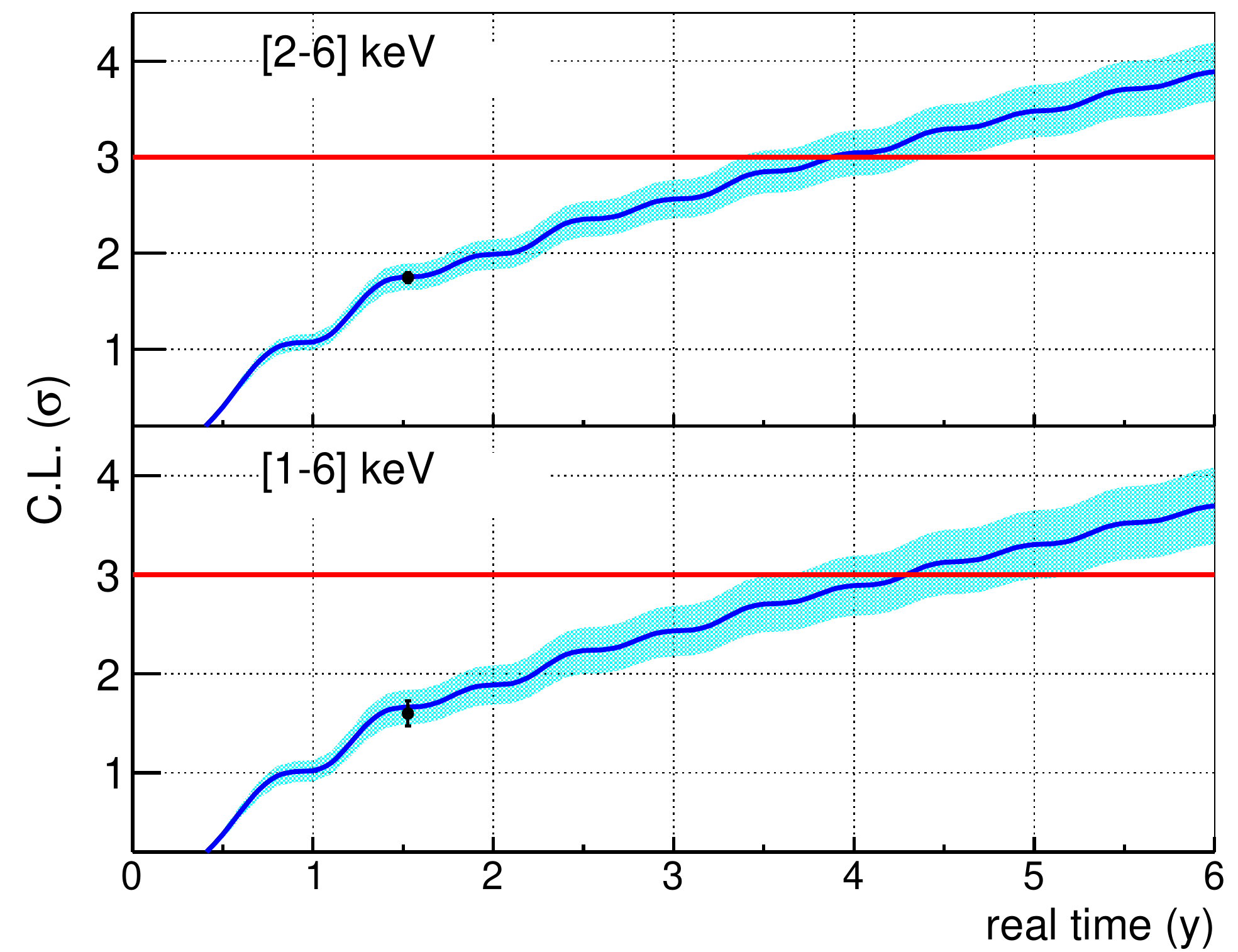}
\caption{
{\anais} sensitivity to the DAMA/LIBRA signal in $\sigma$ C.L. units (see text) as a function of real time in the {\sroi}~keV (upper
panel) and {\roi}~keV (lower panel) energy regions.
The black dots are the sensitivities derived in this work, $\sigma(S_m)$.
The blue bands represent the 68\% C.L. DAMA/LIBRA uncertainty~\cite{Bernabei:2018yyw}.
}
\label{fig:sensitivity}
\end{figure}

The statistical significance of our result is determined by 
the standard deviation of the modulation amplitude distribution, $\sigma(S_m)$,
which would be obtained in a large number of experiments like {\anais} with the present exposure. 
Then, we quote our sensitivity to DAMA/LIBRA result as the ratio $S_m^{DAMA}/\sigma(S_m)$, which directly gives in $\sigma$ units the C.L. at which 
we can test the DAMA/LIBRA signal.
At present, our result  $\sigma(S_m)=0.0058~(0.0063)$~cpd/kg/keV for {\sroi}~keV
({\roi}~keV) corresponds to a sensitivity of 1.75$\sigma$~(1.66$\sigma$) to the DAMA/LIBRA signal. 
In Ref.~\cite{Coarasa:2018qzs} we found an analytical expresion to calculate $\sigma(S_m)$
at a given exposure from the measured background and detection efficiency.
Figure~\ref{fig:sensitivity} (dark blue lines) displays our sensitivity projection calculated following 
Ref.~\cite{Coarasa:2018qzs} 
for the two studied energy 
ranges, whereas the blue bands represent the 68\% uncertainty in $S_m^{DAMA}$ as reported in Ref.~\cite{Bernabei:2018yyw}. 
In the calculation we take into account the {\anais} live time distribution, the background reduction expected
due to decaying isotopes and 
the statistical error in the detection efficiency.
The black dots are the sensitivities derived in this work, including a systematic error estimated by 
changing the time-binning from 1 to 20~days, and considering the systematics in the efficiency~\cite{Amare:2018sxx}. 
The results perfectly agree 
with our estimates, confirming the {\anais} projected sensitivity to the DAMA/LIBRA result. 
A 3$\sigma$ sensitivity should be at reach in 4-5 years of data-taking.
\par
\begin{figure}[b]
\centering
\includegraphics[width=0.50\textwidth]{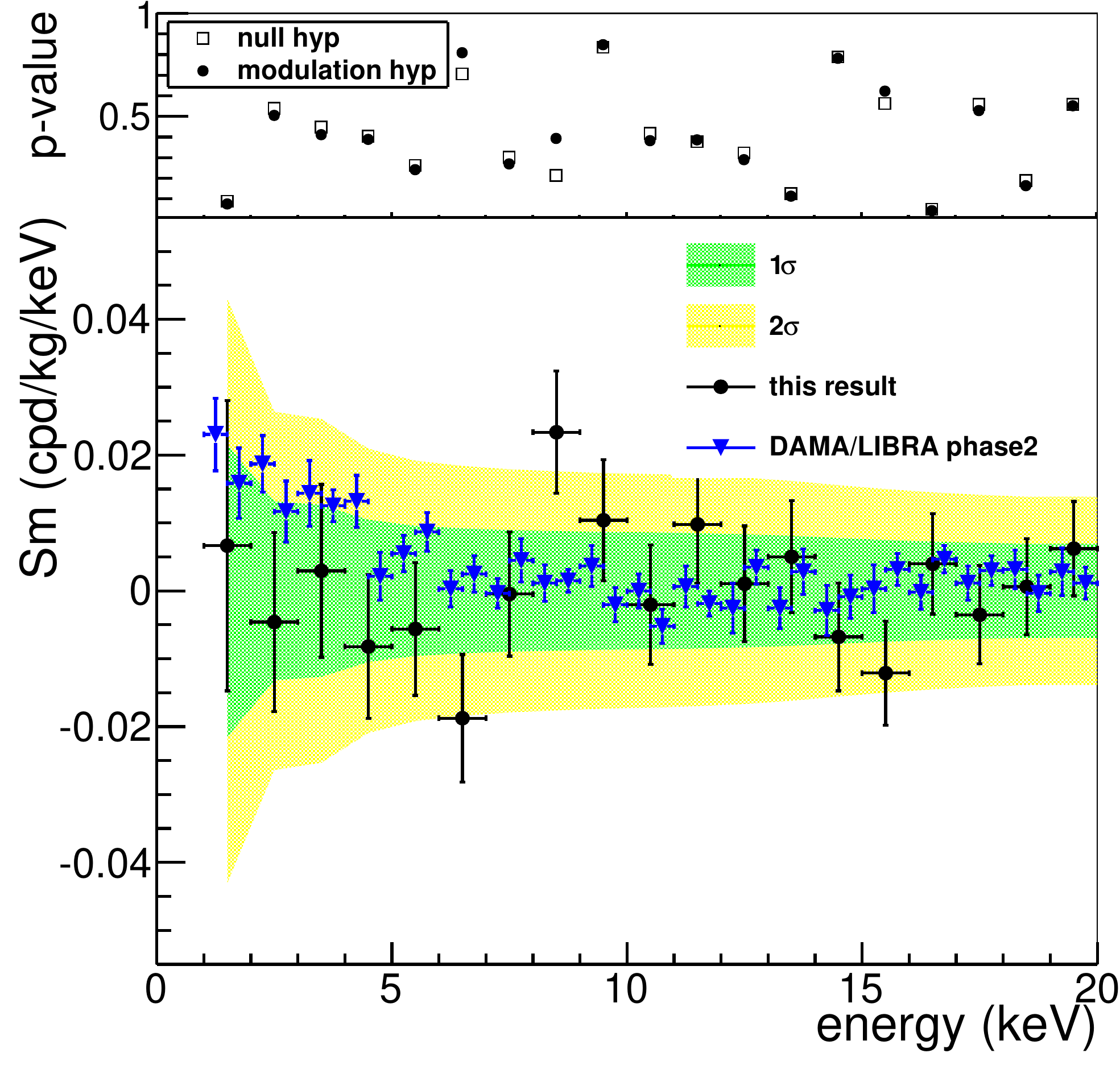}
\caption{Modulation amplitude per 1~keV energy bins combining data from all the modules. 
We show for reference the corresponding DAMA/LIBRA result~\cite{Bernabei:2018yyw}, and the 1$\sigma$ and 2$\sigma$ {\anais} bands following the analysis done in Ref.~\cite{Coarasa:2018qzs}. Top panel compares the p-values of the fits to Equation~\ref{eq:1} with those corresponding to the null hypothesis for every energy bin.}
\label{fig:mod3}
\end{figure}

Finally, Figure~\ref{fig:mod3} presents the best fit amplitudes, $S_m$, calculated 
per 1~keV energy bins, from 1 to 20~keV (bottom panel, black dots), 
together with the DAMA-phase-2 modulation amplitudes 
extracted from Ref.~\cite{Bernabei:2018yyw} (blue triangles).
The top panel shows the p-values for the null (open squares) and 
modulation hypothesis (closed circles) for every energy bin.
All the modulation amplitudes in the ROI are compatible
with 0 and, in general, p-values for the null hypothesis are slightly larger than those
of the modulation hypothesis. The 1$\sigma$ and 2$\sigma$ bands shown in the figure are obtained following Ref.~\cite{Coarasa:2018qzs} for the present {\anais} exposure.
\par
\par
In summary,
to test the DAMA/LIBRA annual modulation result in a model independent way, 
an analysis of the first year and half of data from 
{\anais} experiment has been presented. It is compatible with the 
sensitivity estimates done in Ref.~\cite{Coarasa:2018qzs}, and confirms our goal of reaching
sensitivity at 3$\sigma$ level in five years (see Figure~\ref{fig:sensitivity}) to DAMA/LIBRA result. We derive best fits for 
the modulation amplitude $S_m=-0.0044\pm0.0058$ and $-0.0015\pm0.0063$~cpd/kg/keV, 
in the {\sroi} and {\roi}~keV energy regions, respectively, compatible with the absence of modulation. 
 
\par

\vspace{1cm}

This work has been financially supported by the Spanish Ministerio de 
Econom{\'\i}a y Competitividad and the European Regional Development 
Fund (MINECO-FEDER) under grants No. FPA2014-55986-P and FPA2017-83133-P, 
the Consolider-Ingenio 2010 Programme under grants MultiDark CSD2009-00064 
and CPAN CSD2007-00042 and the Gobierno de Arag{\'o}n and the 
European Social Fund (Group in Nuclear and Astroparticle Physics 
and I.~Coarasa predoctoral grant). We thank the support of the Spanish Red Consolider MultiDark FPA2017-90566-REDC. We acknowledge the technical support from LSC and GIFNA staff. Professor J.A. Villar passed away in August, 2017. Deeply in sorrow, we all thank him for his dedicated work and kindness.

\end{document}